\documentclass[twocolumn,tighten,numberedappendix]{aastex63}
\usepackage{amsmath}
\usepackage{ulem}
\setcitestyle{notesep={ }}

\graphicspath{{./}{figures/}}

\newcommand\alfven{Alfv\'{e}n~}

\def\simlt{\lower.5ex\hbox{$\; \buildrel < \over \sim \;$}}
\def\simgt{\lower.5ex\hbox{$\; \buildrel > \over \sim \;$}}

\def\beq{\begin{equation}}
\def\eeq{\end{equation}}
\def\ba{\begin{eqnarray}}
\def\ea{\end{eqnarray}}
\def\bB{\boldsymbol{B}}
\def\bE{\boldsymbol{E}}

\def\M{{\cal M}}

\def\n{\tilde{n}}
\def\B{{\cal B}}
\def\D{\tilde{\Delta}}

\def\E{{\cal E}}
\def\N{{\cal N}}
\def\I{{\cal I}}

\def\jA{j_{\rm A}}

\def\s{\AB{s}}

\newbox\grsign \setbox\grsign=\hbox{$>$} \newdimen\grdimen \grdimen=\ht\grsign
\newbox\simlessbox \newbox\simgreatbox \newbox\simpropbox
\setbox\simgreatbox=\hbox{\raise.5ex\hbox{$>$}\llap
     {\lower.5ex\hbox{$\sim$}}}\ht1=\grdimen\dp1=0pt
\setbox\simlessbox=\hbox{\raise.5ex\hbox{$<$}\llap
     {\lower.5ex\hbox{$\sim$}}}\ht2=\grdimen\dp2=0pt
\setbox\simpropbox=\hbox{\raise.5ex\hbox{$\propto$}\llap
     {\lower.5ex\hbox{$\sim$}}}\ht2=\grdimen\dp2=0pt
\def\simgt{\mathrel{\copy\simgreatbox}}
\def\simlt{\mathrel{\copy\simlessbox}}

\definecolor{ao(english)}{rgb}{0.0, 0.5, 0.0}
\newcommand{\AB}{\textcolor{blue}}

\begin{document}

\title{
Relativistic Alfv\'en Waves Entering Charge Starvation in the Magnetospheres of Neutron Stars
}

\author[0000-0002-4738-1168]{Alexander Y. Chen}
\affil{JILA, University of Colorado and National Institute of Standards and Technology, 440 UCB, Boulder, CO 80309, USA}

\author[0000-0002-0108-4774]{Yajie Yuan}
\affiliation{Center for Computational Astrophysics, Flatiron Institute, 162 Fifth Avenue, New York, NY 10010, USA}

\author{
Andrei M. Beloborodov
}
\affil{
Physics Department and Columbia Astrophysics Laboratory, Columbia University, 538  West 120th Street New York, NY 10027}
\affil{
Max Planck Institute for Astrophysics, Karl-Schwarzschild-Str. 1, D-85741, Garching, Germany
}

\author[0000-0003-0750-3543]{Xinyu Li}
\affil{Canadian Institute for Theoretical Astrophysics, 60 St George St, Toronto, ON M5R 2M8}
\affil{Perimeter Institute for Theoretical Physics, 31 Caroline Street North, Waterloo, Ontario, Canada, N2L 2Y5}
\correspondingauthor{Alex Chen}
\email{yuran.chen@colorado.edu}

\begin{abstract}
    Instabilities in a neutron star can generate Alfv\'en waves in its
    magnetosphere. Propagation along the curved magnetic field lines strongly
    shears the wave, boosting its electric current $j_{\rm A}$. We derive an
    analytic expression for the evolution of the wave vector $\boldsymbol{k}$
    and the growth of $j_{\rm A}$. In the strongly sheared regime, $j_{\rm A}$
    may exceed the maximum current $j_{0}$ that can be supported by the
    background $e^{\pm}$ plasma. We investigate these ``charge-starved'' waves,
    first using a simplified two-fluid analytic model, then with
    first-principles kinetic simulations. We find that the Alfv\'en wave
    continues to propagate successfully even when
    $\kappa \equiv j_{\rm A}/j_{0} \gg 1$. It sustains $j_{\rm A}$ by
    compressing and advecting the plasma along the magnetic field lines with
    particle Lorentz factors $\sim \kappa^{1/2}$. The simulations show how
    plasma instabilities lead to gradual dissipation of the wave energy, giving
    a dissipation power
    $L_{\rm diss}\sim 10^{35}(\kappa/100)^{1/2} (B_w/10^{11}\,{\rm G})\,\mathrm{erg/s}$,
    where $B_w$ is the wave amplitude. 
    Our results imply that dissipation due to charge starvation is not sufficient to power observed fast radio bursts (FRBs),
    in contrast to recent proposals.
\end{abstract}



\section{Introduction}

Young and active neutron stars can experience quakes which are capable of
launching kHz \alfven waves into the star's magnetosphere
\citep{1989ApJ...343..839B}. This process can power X-ray bursts from magnetars
\citep{1992ApJ...392L...9D}. Quakes have also been associated with glitches in
the rotational frequencies of young radio pulsars
\citep[e.g.][]{1976ApJ...203..213R}, and the quake excitation of magnetospheric
\alfven waves was invoked to explain the chocking of the radio signal from the
Vela pulsar during a glitch \citep{2018Natur.556..219P,2020ApJ...897..173B}.
\alfven waves may also be involved in the production of the fast radio burst
(FRB) detected recently from the galactic magnetar SGR 1935+2154
\citep{2020arXiv200510324T, 2020arXiv200510828B,2020arXiv200506335M}.

The group velocity of an \alfven wave in the magnetosphere of a neutron star is
near the speed of light $c$, and it is directed along the magnetic field
$\boldsymbol{B}_0$. Propagation along the curved magnetic field lines leads to
the growth of $k_\perp$, the wavenumber perpendicular to $\boldsymbol{B}_0$
\citep{2020ApJ...897..173B}. This process can strongly enhance the electric
current in the \alfven wave $\jA\sim (c/4\pi)k_{\perp}B_w$, where $B_w$ is the
wave amplitude.

The maximum current that can be supported by a
plasma with density $n_0$ is $en_0c$, and it is convenient to define the
dimensionless parameter
\begin{equation}
 \kappa \equiv \frac{\jA}{j_0}, \qquad j_0=en_0c.
\end{equation}
The regime of $\kappa>1$ is often called ``charge-starved.'' Charge starvation
has long been invoked as a possible dissipation mechanism of magnetospheric
waves \citep{1989ApJ...343..839B,PhysRevD.57.3219}. Recently, it has been
proposed that waves entering the regime of $\kappa>1$ will develop an enormous
electric field $E_\parallel$ (parallel to $\bB_{0}$) and dissipate a large
fraction of the wave energy, possibly accompanied by strong plasma bunching and
coherent radio emission, which was suggested as a potential mechanism for FRB
emission \citep{2017MNRAS.468.2726K,2020MNRAS.494.2385K,2020arXiv200506736L}.

In this paper, we examine the behavior of charge-starved \alfven waves. In
particular, we wish to know what $E_\parallel$ is induced, how much of the wave
energy is dissipated, what plasma instabilities will arise, and what is the
resulting particle distribution. We begin with a discussion of how an \alfven
wave packet can become charge-starved as it propagates in the magnetosphere of a
neutron star (Section~\ref{sec:shear}). Then we investigate what happens with
the wave as it enters the regime of $\kappa>1$. We first use a simplified
analytical model (Section~\ref{sec:two-fluid}), then perform direct kinetic
simulations of the plasma dynamics in the wave (Section~\ref{sec:sim}).

\section{Propagation and Shearing of an Alfv\'en Wave Packet}

\label{sec:shear}

Consider an \alfven wave packet launched into the magnetosphere by a shear
motion of the neutron star crust. Let $B_w$ be the wave amplitude and
$\ell_\perp$ be the perpendicular size of the packet. The initial $\ell_\perp$
equals the size of the sheared region of the stellar surface, which determines
the current density in the packet $\jA\sim (c/4\pi) B_w/\ell_\perp$ that flows
along the background magnetic field $\boldsymbol{B}_{0}$. Small amplitude waves
with $B_{w}\ll B_0$ will propagate along the magnetic field lines without
disrupting the structure of the magnetosphere. Below we examine two main effects
that will affect the evolution of $j_{A}$ as the wave packet propagates away
from the star.

First, the divergence of the dipole field lines will increase $\ell_{\perp}$ and
decrease $B_w$. The distance between magnetic flux surfaces increases with
radius $r$ as $\ell_{\perp} \propto r^{3/2}$ while $B_w\propto r^{-3/2}$. This
effect leads to a scaling of $\jA\propto r^{-3}$. Incidentally, the
Goldreich-Julian charge density $\rho_\mathrm{GJ}$ in a rotating dipole
magnetosphere also decreases as $r^{-3}$ \citep{1969ApJ...157..869G}. This led
\cite{2020MNRAS.494.2385K} to suggest that in a magnetosphere with plasma
density $n=\M \rho_{\rm GJ}/e$, the \alfven wave can become charge-starved if
the multiplicity $\M$ decreases with radius. On closed field lines, when the
wave packet reaches the magnetic equator and turns back to the star, it follows
the converging field lines and $\ell_{\perp}$ decreases again. If the
divergence/convergence of the magnetic flux surfaces were the only effect,
$j_{A}$ would come to its original value when reaching the stellar surface in
the opposite hemisphere.

However, there is a second effect that can enhance the perpendicular gradient of
the \alfven wave packet, hence increasing $j_{A}$, especially on closed field
lines. Different dipole field lines have different lengths, and the parts of the
packet propagating along the longer field lines lag behind, leading to a strong
shear of the packet. \citet{2020ApJ...897..173B} described this effect as
``de-phasing'', and studied the evolution of $j_{A}$ as the wave keeps bouncing
in the closed magnetosphere. Here we use a different approach to compute the
evolution of $k_{\perp}$ within a single pass in the magnetosphere.

A dipole magnetic field line is parametrized by $r = r_{m}\sin^{2}\theta$, where
$r_{m}$ is the radius where it crosses the magnetic equator $(\theta = \pi/2)$.
The field line starts at the stellar surface (radius $r_\star$) at polar angle
$\theta_0$ related to $r_m$ by $\sin^2\theta_0=r_\star/r_m$. It is convenient to
use variable $\mu=\cos\theta$, which varies along the closed field line between
$\mu_0$ and $-\mu_0$. Starting from the northern
footpoint $\mu_0$
one can integrate the length 
along the dipole field line to a given point $\mu$,
\begin{align}
  s(r_m, \mu)
  &
    =r_{m}\int_{\mu}^{\mu_0}\sqrt{3\mu_1^2+1}\,d\mu_1 \nonumber\\
  &= r_m [F(\mu_0) -F(\mu)],
\label{eq:s}
\end{align}
where 
\begin{equation}
    \label{eq:F}
    F(\mu) = \frac{1}{2}\mu \sqrt{1 + 3\mu^{2}} + \frac{\sinh^{-1}(\sqrt{3}\mu)}{2\sqrt{3}}.
\end{equation}
Note that $\mu_0=(1-r_\star/r_m)^{1/2}$, so $F(\mu_0)$ is a function of $r_m$.

Let us now consider an \alfven wave with frequency $\omega$ launched from the stellar surface in the northern hemisphere.
The wave at $t>0$ is described by $B_w(r/r_{*})^{-3/2}\exp[i\Phi(t,\boldsymbol{r})]$, where
\begin{equation}
  \Phi(t,\boldsymbol{r})=
\Phi_0(r_m)
-\omega t+k_\parallel s(\boldsymbol{r}), \qquad k_\parallel=\frac{\omega}{c}.
\end{equation}
As long as $c/\omega\ll r_{*}$ and the initial $\ell_{\perp} \ll r_{*}$, the
wavevector is $\boldsymbol{k} = \nabla\Phi = \nabla\Phi_0 + (\omega/c)\nabla s$.
The first term is the contribution to $\boldsymbol{k}$ from the initial profile
of the perturbation. The evolution of $\nabla_\perp\Phi_0$ follows the
divergence of the field lines, and is essentially the first effect we discussed
above. We are interested in the evolution of $k_\perp$ due to the second term,
$k_\perp = (\omega/c)\nabla_\perp s$.

To evaluate $\nabla_{\perp}s$, it is convenient to introduce the dipole
coordinates following \citet{2006physics...6044S}:
\begin{equation}
    \label{eq:dipole-coord}
    \eta = \frac{r}{\sin^{2}\theta} = r_{m},\quad \chi = \frac{\cos\theta}{r^{2}} = \frac{\mu}{r^{2}}.
\end{equation}
The $\eta$-coordinate coincides with $r_m$, and uniquely labels the field
  lines, while the $\chi$-coordinate varies along a single field line. This is
an orthogonal coordinate system with metric elements:
\begin{equation}
    h_{\eta} = \sqrt{g_{\eta\eta}} = \frac{\sin^{3}\theta}{\sqrt{1 + 3\mu^{2}}},\quad h_{\chi} = \sqrt{g_{\chi\chi}} = \frac{r^{3}}{\sqrt{1 + 3\mu^{2}}}.
\end{equation}
The metric coefficient $h_{\eta}$ quantifies the distance between the dipole
flux surfaces. The perpendicular derivative is simply given by:
\begin{equation}
    \frac{ck_{\perp}}{\omega} = \nabla_{\perp}s = \frac{1}{h_{\eta}}\left.\frac{\partial s}{\partial \eta}\right|_{\chi},
\end{equation}
which can be evaluated to be:
\begin{equation}
    \begin{split}
        \frac{ck_{\perp}}{\omega} = &\frac{\sqrt{1+3\mu^2}}{\sin^3\theta}\bigg([F(\mu_0)-F(\mu)]  \\
        & +\left.\frac{\sqrt{1+3\mu_0^2}}{2\mu_0}\,\frac{r_\star}{r_{m}} - 2\frac{\mu\sin^{2}\theta}{\sqrt{1 + 3\mu^{2}}}\right).
    \end{split}
\end{equation}

\begin{figure}[h]
    \centering
    \includegraphics[width=0.47\textwidth]
    {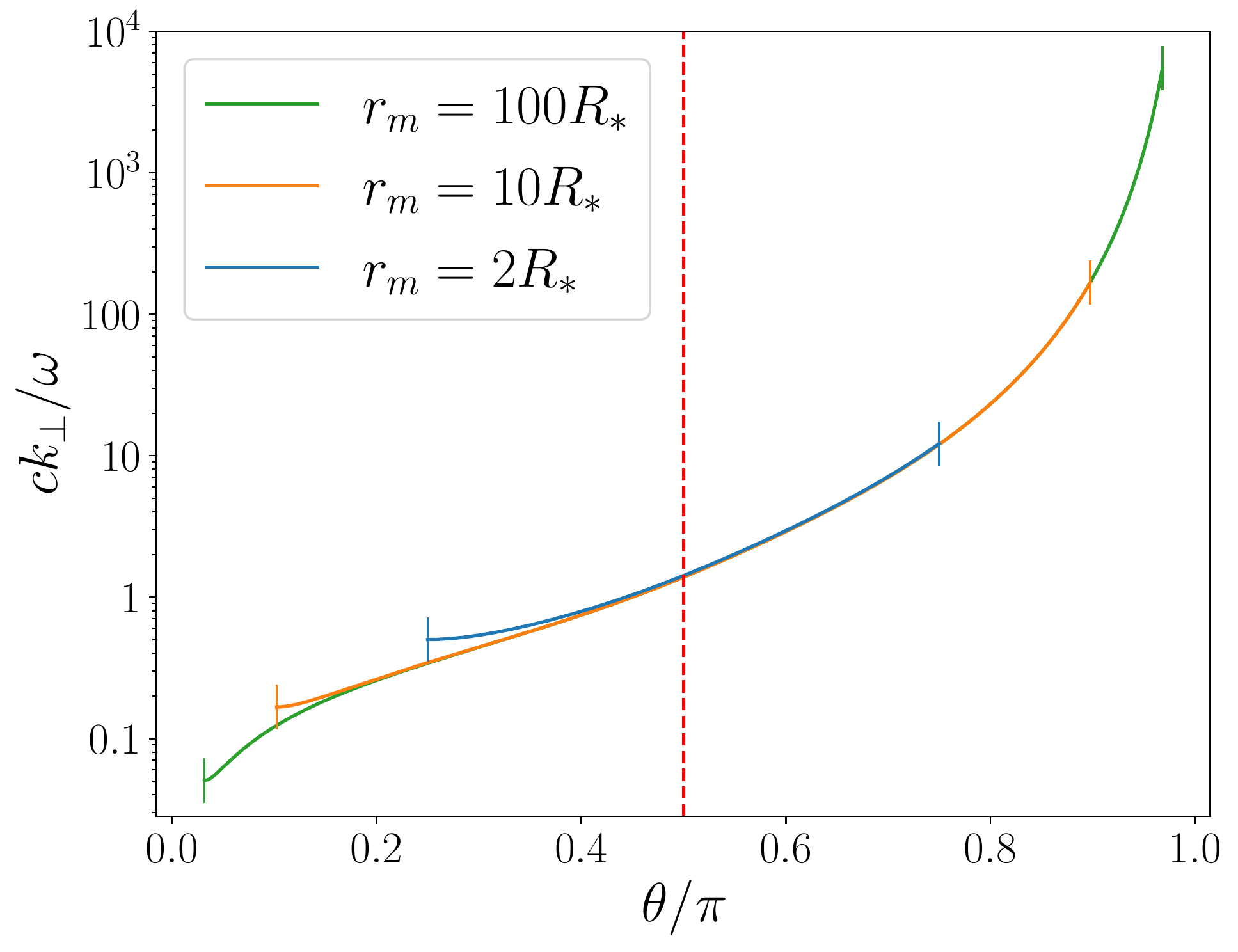}
    \caption{The evolution of $k_\perp$ on field lines of different $r_m$. The
      vertical markers indicate the footpoints of the magnetospheric field line
      on the star. The red dashed vertical line shows the magnetic equator
      $\theta = \pi/2$.}
    \label{fig:shearing-func}
\end{figure}

The variation of $k_{\perp}$ along a given field line is shown in
Figure~\ref{fig:shearing-func}. For extended field lines, $r_{m}\gg r_{*}$,
$k_{\perp}$ can grow to very large values in the southern hemisphere. This is a
combination of two effects. The cumulative path difference
$\partial s/\partial \eta$ grows fastest near the equator, which leads to
$k_{\perp} \sim \omega/c$. Then, due to the field line convergence in the
southern hemisphere, the existing $k_{\perp}$ is increased by a factor of
$1/h_{\eta} \sim (r_{m}/r_{*})^{3/2}$ when the wave reaches the southern
footpoint. After $N$ consecutive bounces, the wave will accumulate a total
$ck_{\perp} \propto N\omega/h_\eta$, 
consistent with the result of \citet{2020ApJ...897..173B} (their equation~(38)
used the approximation $\nabla_\perp\sim r^{-1}\partial_\theta$, which is valid
for $\theta$ away from the equatorial plane).

The evolution of $ck_{\perp}/\omega$ results in the \alfven wavefront becoming
increasingly oblique with respect to the background field $\boldsymbol{B}_{0}$.
The angle $\psi$ between the wave vector $\boldsymbol{k}$ and
$\boldsymbol{B}_{0}$ grows as
$\tan\psi=k_\perp/k_\parallel = ck_{\perp}/\omega$. The dimensionless parameter
$\kappa$ grows as:
\begin{equation}
    \kappa = \frac{j_{A}}{en_{0}c}\sim \frac{\omega_B\omega \tan\psi}{\omega_p^2}\,\frac{B_w}{B_0},
\end{equation}
where $\omega_{p} = \sqrt{4\pi e^{2}n_{0}/m_{e}}$ is the background plasma
frequency, and $\omega_{B}=eB_0/m_ec$ is the gyro-frequency of electrons in the
background magnetic field. Due to the ever growing $\tan\psi$, the effect of
wave shearing can eventually lead to charge-starvation, $\kappa > 1$, even for
waves of modest amplitudes.

\section{Plasma dynamics in the wave: two-fluid model}

\label{sec:two-fluid}

\subsection{Problem Setup}

As a first step toward understanding charge-starved \alfven waves, we examine a
simple two-fluid model of plasma motion in a plane wave propagating into a
uniform background. The uniform approximation is reasonable for sufficiently
short waves. We take the background to be a cold neutral $e^\pm$ plasma with
density $n_{+} = n_{-} = n_{0}=\mathrm{const}$ immersed in a uniform magnetic
field $\bB_0$. A gradual change of $\kappa$ may then be treated as an adiabatic
effect on the quasi-steady plane wave.

We are interested in the highly magnetized regime with
$\sigma\equiv B_0^2/4\pi n_0 m_e c^2 \gg 1$, so that the propagation speed of
the wave along $\bB_0$ nearly equals $c$. The electromagnetic field of a
steadily propagating wave is then only a function of $t-s/c$, where the
$s$-coordinate runs along $\bB_0$. Let the $x$-axis be along the wave vector
$\boldsymbol{k}$, the wave magnetic field $\boldsymbol{B}$ depends on
\begin{equation}
  \xi=t-\frac{s}{c}=t-\frac{x}{c\cos\psi}.
\end{equation}
The propagation speed along $x$ is $V_x=c\cos\psi$. The wave electric field $\bE$ is
related to the magnetic field $\bB$ by
\begin{equation}
    \label{eq:Ewave}
  \bE(\xi)=\bB(\xi)\times\frac{\boldsymbol{B}_0}{B_0}.
\end{equation}
The wave fields $\bB$ and $\bE$ are both perpendicular to the background field.
The electric current density in the wave is parallel to $\bB_0$ and its value is
\begin{equation}
\label{eq:j}
  j_{A}(\xi)=\frac{\tan\psi}{4\pi}\frac{dB}{d\xi}.
\end{equation}
The charge density in the wave, $\rho=\nabla\cdot\bE/4\pi$ satisfies the
relation
\begin{equation}
  c\rho(\xi)=j_{A}(\xi).
\end{equation}

The above description gives an exact MHD solution in the force-free limit
$\sigma\rightarrow\infty$ with no charge starvation. It relies on the implicit
assumption that there is always enough plasma to conduct the required electric
current $j_{A}$. We will next examine the dynamics of the $e^{\pm}$ particles
for any given background plasma density $n_{0}$, especially in the regime
$\kappa > 1$, when the assumption of copious plasma supply may not be valid.

The characteristic gyro-frequency $\omega_B$ in the neutron star
magnetosphere is many orders of magnitude greater than $\omega$. Therefore,
$e^{\pm}$ particles in the wave move with velocities $\boldsymbol{v}$ along the
magnetic field lines, like beads on a wire. Only $E_{\parallel}$ is relevant for
their dynamics. For small amplitude waves $B_{w}\ll B_{0}$, the field lines are
bent only by a small angle, and
$v_\parallel =\boldsymbol{v}\cdot \bB_0/B_0 +{\cal O}(B_w^2/B_0^2)$. Therefore
we approximate the particle motion as parallel to $\boldsymbol{B}_{0}$.

In order to conduct the required electric current, an $E_{\parallel}$ will be
induced to accelerate the electrons and positrons in opposite directions,
creating two plasma streams. 

\subsection{Two-fluid Model}

We first examine a simple model assuming that the $e^\pm$ streams remain cold,
neglecting any possible instabilities that may arise. This ``two-fluid'' model
captures some basic features of the plasma dynamics in the wave. The parallel
electric field regulating the velocities of the $e^\pm$ streams is
non-dissipative in the two-fluid model, as the particles will come to rest
behind the wave.

The two cold fluids are described by their densities $n_\pm$ and velocities
$v_\pm$. Both are functions of $\xi$ in a steadily propagating wave.
Their values in the background plasma, $n_\pm=n_0$ and $v_\pm=0$, give the
boundary conditions ahead of the wave for the profiles $n_\pm(\xi)$ and
$v_\pm(\xi)$.
The density and velocity of each stream satisfies the continuity equation
$\partial_tn_\pm+\partial_x(n_\pm v_{\pm,x})=0$, where
$v_{\pm,x}=v_\pm\cos\psi$. This gives
\begin{equation}
    \label{eq:continuity}
    \frac{d n_{\pm}}{d\xi} - \frac{d}{d\xi}(n_{\pm}\beta_{\pm}) = 0,
\end{equation}
where $\beta_\pm=v_\pm/c$. One then finds $(1-\beta_\pm)dn_\pm=n_\pm d\beta_\pm$ and
\begin{equation}
\label{eq:npm}
    n_{\pm}(\xi) = \frac{n_{0}}{1 - \beta_{\pm}(\xi)},
\end{equation}
where we used the boundary condition ahead of the wave: $n_\pm=n_0$ when
$v_\pm=0$. In the two-fluid picture, the continuity equation automatically
implies the relation $j = \rho c$, where $j=e(n_+v_+-n_-v_-)$ and
$\rho=e(n_+-n_-)$.

The fluid velocities are related to
$\kappa = j_{A}/en_{0}c$ by
\begin{equation}
\label{eq:kappa}
  \frac{1}{1-\beta_+}-\frac{1}{1-\beta_-}=\kappa(\xi).
\end{equation}
In a successfully propagating \alfven wave, the plasma motion must sustain
$j=\jA$ given in Equation~(\ref{eq:j}), which determines $\kappa(\xi)$. Our goal
is to find $\beta_\pm(\xi)$ under a given $\kappa(\xi)$, and then check what
happens in the charge starvation regime of $\kappa>1$.

Particles are governed by the equation of motion $dp_\pm/dt=\pm e E_\parallel$,
where $p_\pm=\gamma_\pm m_ev_\pm$, $\gamma_\pm=(1-\beta_\pm^2)^{-1/2}$, and
$d/dt=\partial_t+v_\pm d/ds=(1-\beta_\pm)d/d\xi$. This equation gives
\begin{equation}
    \label{eq:accel}
    \frac{dq_{\pm}}{d\xi}=\pm \frac{eE_{\parallel}}{m_ec}, \qquad
    q_\pm\equiv \gamma_\pm(1-\beta_\pm).
\end{equation}
It implies $d(q_++q_-)/d\xi=0$. Then using the boundary condition $q_+=q_-=1$
ahead of the wave, we obtain
\begin{equation}
\label{eq:1}
   q_+ + q_-=2.
\end{equation}
Rewriting Equation~(\ref{eq:kappa}) in terms of $q_\pm$,
\begin{equation}
\label{eq:2}
  \frac{1}{q_+^2}-\frac{1}{q_-^2}=2\kappa(\xi),
\end{equation}
we obtain two equations for $q_\pm$, which can be easily solved for any given
$\kappa(\xi)$. Once $q_\pm(\xi)$ are found, we also obtain
$eE_\parallel=m_ec\, dq_+/d\xi$.

\begin{figure}[t]
    \centering
    \includegraphics[width=0.47\textwidth]{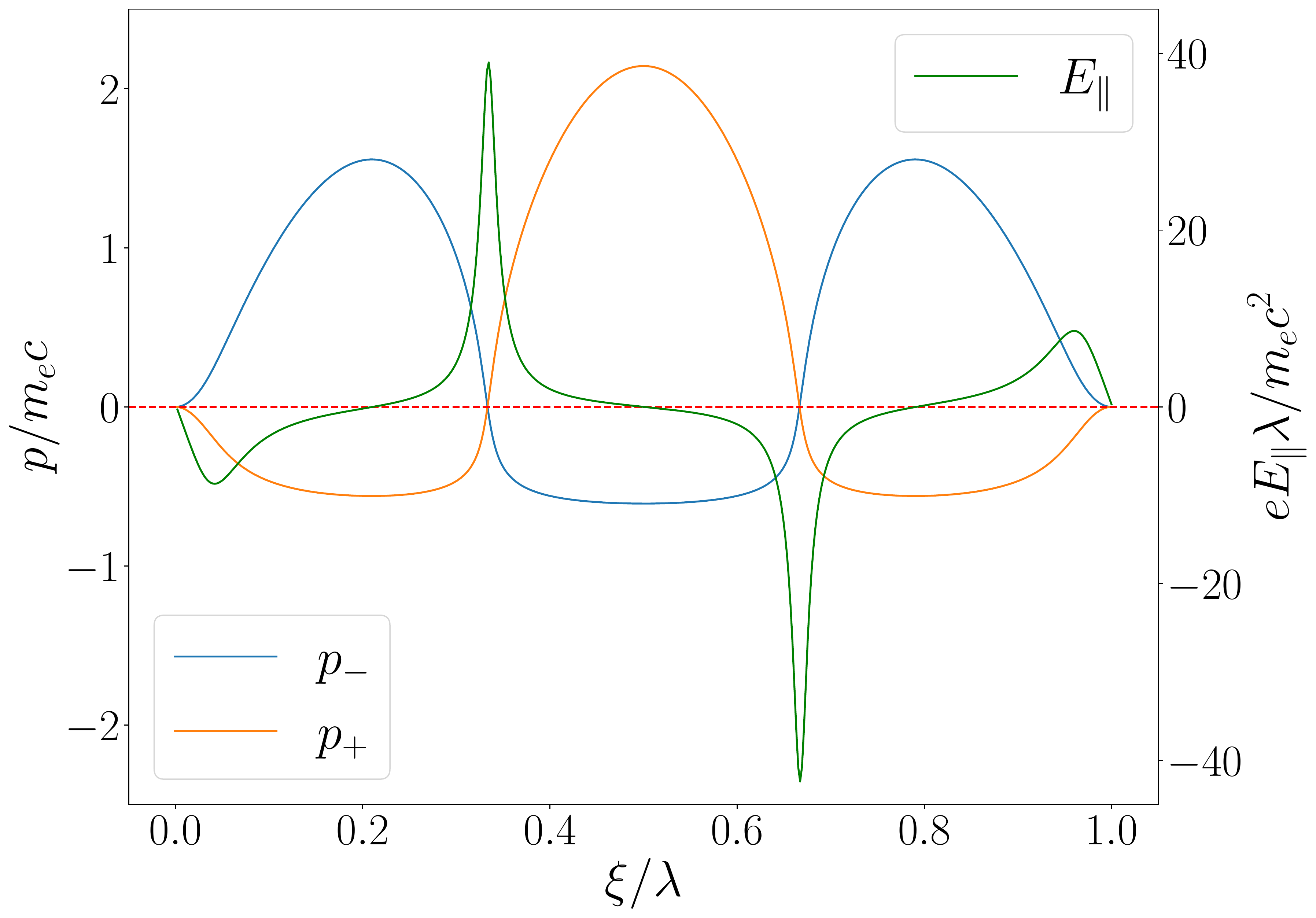}
    \caption{ The momentum profile $p_\pm(\xi)$ of the electron (blue) and
      positron (orange) streams in the two-fluid model, for the wave profile
      given in equation~\eqref{eq:profile}. The corresponding $E_\parallel(\xi)$
      is shown by the green curve. The maximum $\kappa$ in this example is $10$,
      reached at the center of the wave profile. }
    \label{fig:2fluid_solution}
\end{figure}

A well-behaved solution to the Equations~(\ref{eq:1}) and (\ref{eq:2}) exists
for both $|\kappa|<1$ and $|\kappa|>1$. In particular, consider $\kappa\gg 1$,
the strongly ``charge-starved'' regime with $j>0$. Then the solution is
$q_+^{-2}\approx 2\kappa\gg q_-^{-2}\approx 1/4$. Using the relation
$q^{-1}=\gamma(1+\beta)$, we find \beq \gamma_+\approx
\left(\frac{\kappa}{2}\right)^{1/2}, \quad \gamma_-\approx \frac{5}{4} \qquad
(\kappa\gg 1). \eeq Figure~\ref{fig:2fluid_solution} shows the solution for
$p_\pm(\xi)$, and the corresponding $E_\parallel(\xi)$, for a plane wave with
the following profile
\begin{equation}
    \label{eq:profile}
    \frac{B}{B_w}=
\begin{cases}
        \displaystyle
        \frac{8}{3\sqrt{3}}\sin\left( 2\pi\frac{\xi}{\lambda} \right)\sin^{2}\left(\pi\frac{\xi}{\lambda}
        \right),&\quad 0<\xi<\lambda \\
        \displaystyle 0,&\quad \mathrm{otherwise}
    \end{cases}
\end{equation}
This profile describes an isolated sine pulse, with the additional factor of
$\sin^2(\pi\xi/\lambda)$ introduced to make the derivative $dB/d\xi$ smoothly
vanish at the boundaries of the pulse $\xi=0,\lambda$. In our example,
$\kappa(\xi)$ reaches the maximum $\kappa_w = 10$ at $\xi=0.5\lambda$.

The wave achieves the required $j$ and $\rho$ by inducing $E_\parallel$ that
sweeps the $e^\pm$ particles along $\bB_0$. This sweeping compresses the two
fluids by the factors of $(1-\beta_\pm)^{-1}$ (Equation~\ref{eq:npm}). In the
regime of $\kappa\gg 1$ the compression factor is large for $e^+$,
$(1-\beta_+)^{-1}\approx 2\gamma_+^2$. Developing
$\gamma_\pm\approx (\kappa/2)^{1/2}$ is sufficient to enhance the local density
of $e^+$ by the factor of $\kappa$ and thus achieve $\rho$ and $j$ required by
the wave. Particles move through the wave with the relative speed
$c-v\approx c/2\gamma^2$ when $\gamma\gg 1$; therefore it takes time
$t\sim \kappa \lambda/c$ for the plasma to cross the wave.

It is convenient to define
\begin{equation}
  n_w=\frac{\jA}{ec}.
\end{equation}
In the charge-starved regime $\kappa\gg 1$, the wave carries the density
$n\approx n_w$, and the electromagnetic energy available per particle is
described by
\begin{equation}
    \sigma_{w} = \frac{B_{w}^{2}}{4\pi n_{w}m_{e}c^{2}}.
\end{equation}
The wave propagation is weakly affected by charge-starvation as long as
$\gamma_{+} \approx (\kappa/2)^{1/2}\ll \sigma_w$. Otherwise the plasma kinetic
energy will become comparable to the field energy, leading to significant
deviations of $\bE(\xi)$ and $\bB(\xi)$ from the force-free solution.

In neutron star magnetospheres, the plasma frequency $\omega_{p}$ is much higher
than the frequencies of \alfven waves launched by crustal motion. This implies
that the two-fluid model is deficient, because the cold two-stream configuration
is unstable on the short plasma timescale $\sim \omega_p^{-1}$. In particular,
consider the vicinity of a maximum of $p_\pm$ in
Figure~\ref{fig:2fluid_solution}, where the two-fluid model gives
$E_\parallel\approx 0$. Since the parameters of the $e^\pm$ streams are varying
slowly compared to the plasma scale, $\omega\ll \omega_{p}$, one can use the
standard linear instability analysis \citep[e.g.][]{1986islp.book.....M} to find
that the most unstable mode is near $k \sim \omega_{p}/c$ with growth rate
$\Gamma \sim \omega_{p}$. The instability will heat the plasma streams and mix
them in the phase space. It is difficult to predict analytically the
consequences of the nonlinear saturation of the instability. Therefore, we
employ direct kinetic plasma simulations to find a self-consistent solution.

\section{Numerical Simulations}

\label{sec:sim}

\begin{figure*}[t]
    \centering
    \includegraphics[width=0.95\textwidth]{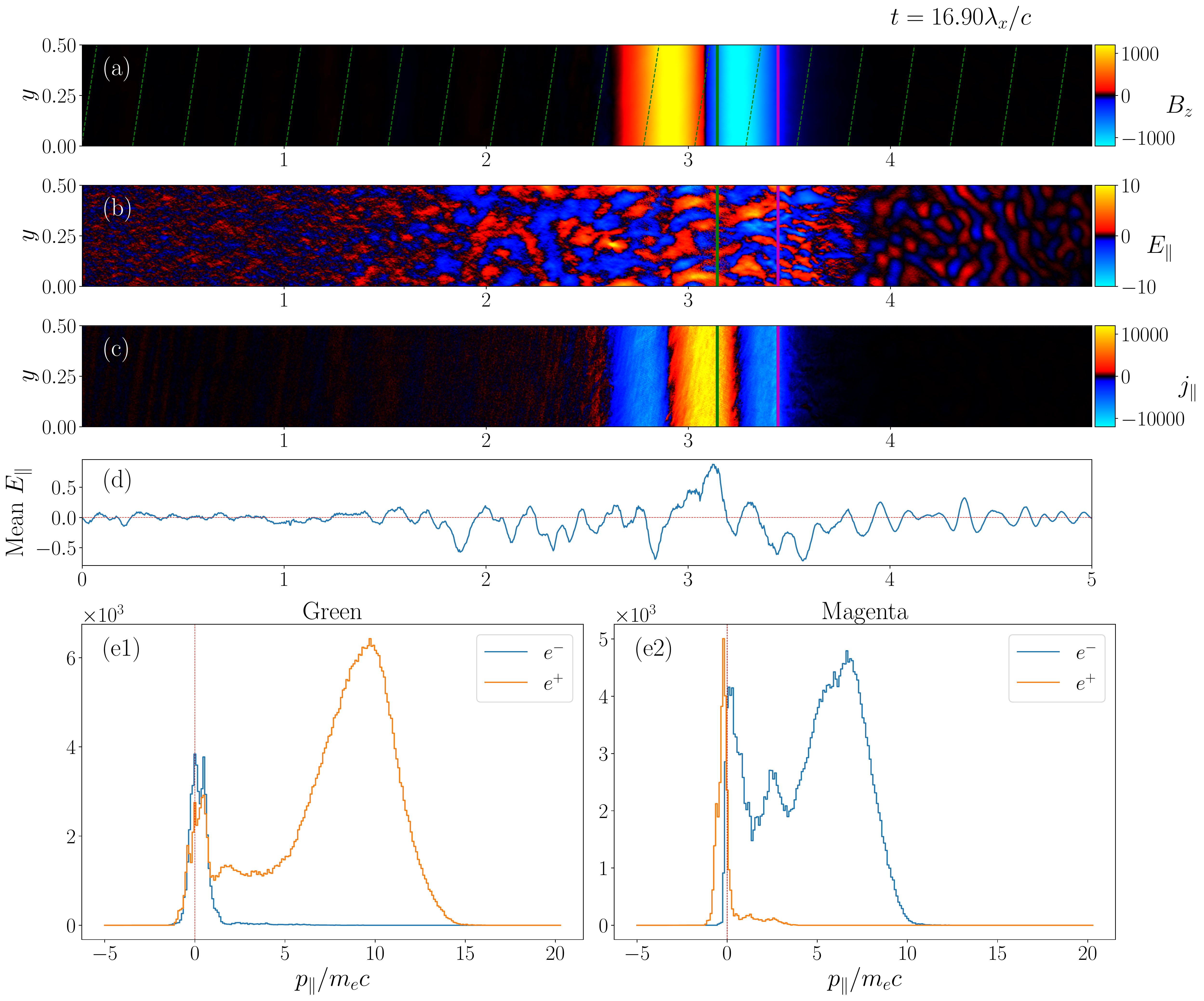}
    \caption{A snapshot of the simulation with $\kappa = 12$. Panels from top to bottom are: (a)
      the wave magnetic field $B_{z}$ as color plot and $B_{0}$ as dashed green
      lines; (b) parallel electric field; (c) parallel current density; (d)
      $E_{\parallel}$ averaged over $y$; (e1) momentum distribution of $e^{\pm}$
      in the green region depicted in panels (a)--(c); (e2) momentum
      distribution of $e^{\pm}$ in the magenta region in panels (a)--(c). $B$
      and $E$ are measured in units of $m_{e}c^{2}/e\lambda_{x}$, while $j$ is
      measured in units of $m_{e}c^{3}/4\pi e\lambda_{x}^{2}$. These are the
      units used in the code.}
    \label{fig:simulation}
\end{figure*}

\subsection{Simulation Setup}

We set up a series of Particle-in-Cell (PIC) simulations using our own GPU-based
PIC code \emph{Aperture}\footnote{\url{https://github.com/fizban007/Aperture4}}.
We use a two-dimensional, elongated Cartesian box with periodic boundary
conditions in the $y$ direction. An \alfven wave is initialized at the left end
of the box with the profile described by Equation~\eqref{eq:profile}, with
magnitude $B_{w}$ and wave $\boldsymbol{B}$ pointing in the $z$ direction. The
wave electric field is initialized using equation~\eqref{eq:Ewave}. The
background magnetic field $\boldsymbol{B}_{0}$ is inclined with respect to the
$x$-axis by an angle $\psi$. In our simulations,
$\cos\psi = \hat{\boldsymbol{B}}_{0}\cdot \hat{\boldsymbol{x}} = 0.15$. As the
wave propagates along $\boldsymbol{B}_{0}$, it will move in the box along the
$x$ direction. The effective length of propagation is much longer than the box
length due to the inclination of the background field. A damping layer is placed
at the end of the box $x = L_{x}$ to prevent the reflection of any plasma waves.

\begin{figure*}[t]
    \centering
    \plottwo{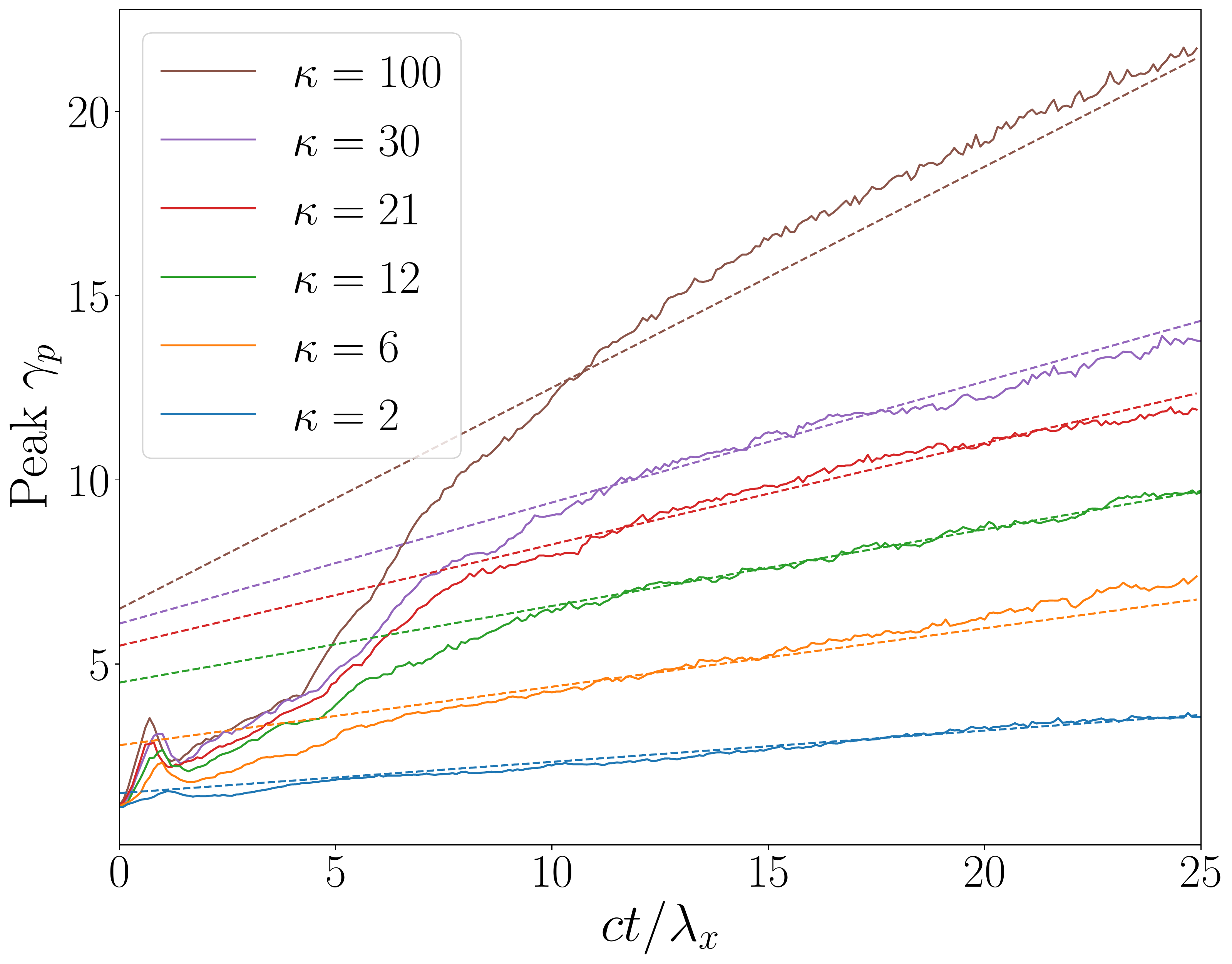}{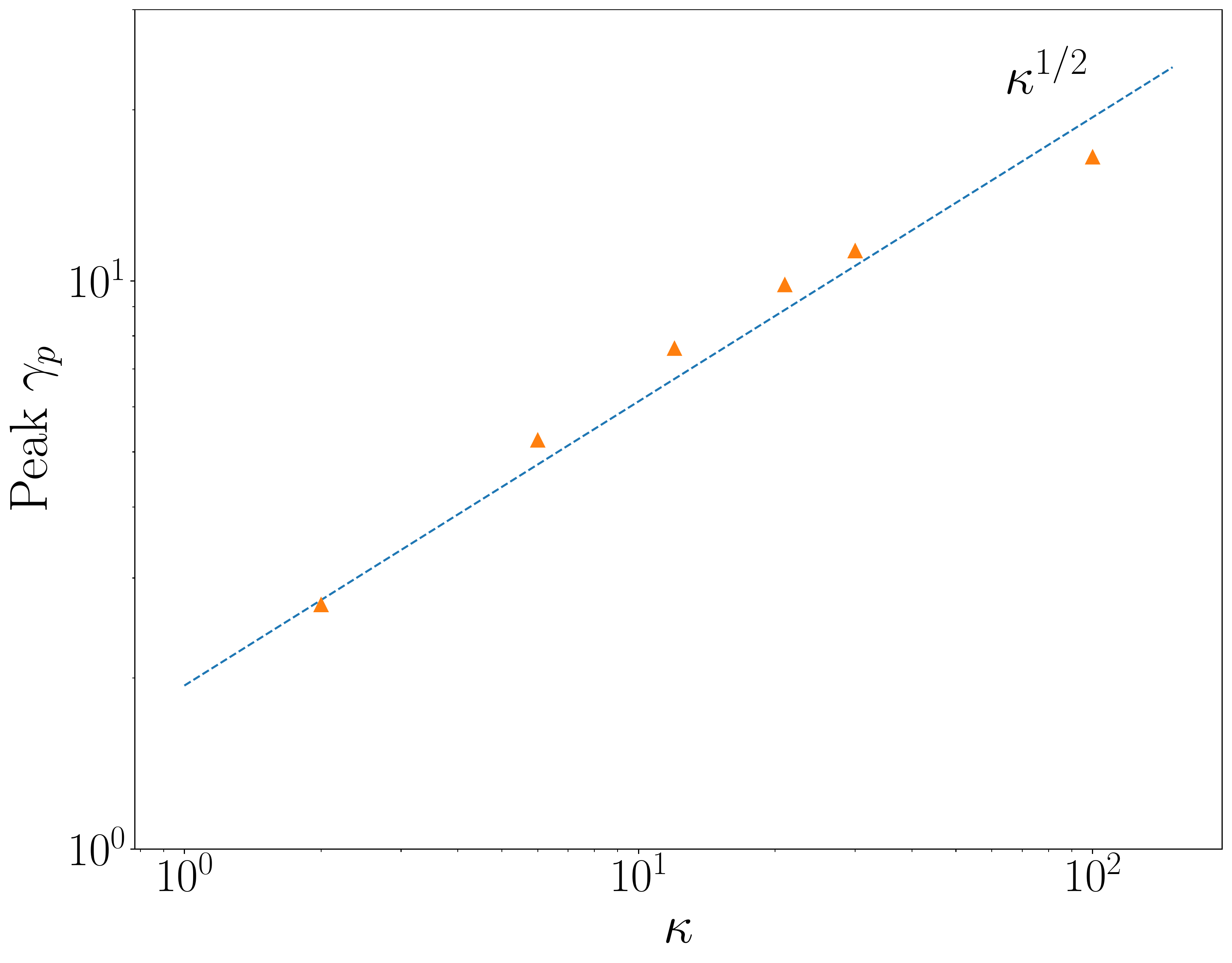}
    \caption{Particle Lorentz factor vs $\kappa$. Left panel: growth of peak
      Lorentz factor of the current conducting particles in the wave at
      different $\kappa$. Dashed lines have slopes $0.4\sqrt{\kappa}/\lambda$,
        and are acceleration models with constant $E_{\parallel}$. Right panel:
      orange triangles are peak Lorentz factors at $ct/\lambda_{x} = 15$, while
      the blue dashed line indicates a simple power law
      $\gamma_{p}\propto \kappa^{1/2}$, which is the prediction of the two-fluid
      toy model.}
    \label{fig:kappa}
\end{figure*}

We start with a small amplitude wave, with $B_{w}/B_{0} = 0.1$. Inside the wave
packet, we initialize a pair plasma that satisfies $\rho = j_{A}/c$ and
$j = j_{A}$ with a small initial multiplicity, $n_{+} + n_{-} = 3j_{A}/ec$. The
space outside the wave packet is filled with a low density neutral plasma with
$n_{+} = n_{-} = n_{0}$. We typically have 5--10 particles per cell
corresponding to $n_{0}$. Depending on the value of $\kappa$ the number of
particles per cell in the wave is often much larger. The characteristic plasma
skin depth $c/\omega_{p}$ in the wave is set by $j_{A}/ec$, and is typically
$\sim 1/200$ of the wavelength in the $x$ direction, $\lambda_{x}$. The box
size is $L_{x} = 5\lambda_{x} = 10L_y$, and has a total resolution $5120\times 512$.
This translates to $\sim 5$ cells per plasma skin depth. Outside the wave
packet, where plasma density $n_{0} \ll j_{A}/ec$, the plasma skin depth is
resolved with many more cells.

There are two dimensionless parameters, $\kappa$ and $\sigma_{w}$, that govern
the physics in this problem. In the following discussion we refer to $\kappa$ as
its maximum value at the center of the wave profile. In the simulations shown
below we always keep $\sigma_{w} \sim 100$ so that
$\sigma_{w}\gg \sqrt{\kappa}$. This is the realistic regime for \alfven waves in
a neutron star magnetosphere. The simulation results will verify that, in this
limit, the amount of wave energy converted to particle kinetic energy is small,
and the wave electromagnetic fields remain close to the initial force-free
solution.

\subsection{Waves in A Uniform Background}

\label{sec:const-n0}

We performed a series of simulations where $\kappa$ is constant in the box, and
increase $\kappa$ from 2 to 100 between the different runs. In these
simulations, we observe that the cold two-stream instability rapidly develops
and heats up the plasma streams. As a result of the nonlinear saturation of the
instability, $E_{\parallel}$ breaks into Langmuir modes that are advected with
the \alfven wave. Figure~\ref{fig:simulation} shows a snapshot of such a state
in one of our PIC simulations with $\kappa = 12$. The plasma momentum
distribution in the wave can be generally described as a mildly relativistic,
current conducting beam traveling into a neutral plasma at rest. This
configuration is also subject to a warm version of the two-stream instability,
exciting electrostatic wave modes that scatter the fast moving particles to
lower velocities, creating a ``bridge'' between the two momentum peaks (see
panel (e) in Figure~\ref{fig:simulation}). These slower particles tend to fall
behind the current-carrying beam, reducing $j$. As a result, the plasma responds
by inducing a small $E_{\parallel}$ on average which keeps accelerating the
particles traveling in the wave. Effectively, this creates an anomalous
resistivity that
continually dissipates the \alfven wave energy. A small fraction of this
dissipated energy is converted into plasma waves launched into the upstream, but
most of the energy goes into gradual acceleration of the relativistic beam.

Figure~\ref{fig:kappa} shows this gradual and continual acceleration of the
current conducting particles, as well as the scaling of their Lorentz factor
with $\kappa$. At any given time, the peak Lorentz factor scales as
$\sqrt{\kappa}$, which agrees with the two-fluid toy model. The acceleration in
all cases seems to be consistent with an average
$E_{\parallel} \approx 0.4\sqrt{\kappa}m_{e}c^{2}/e\lambda$, where $\lambda$ is
the full wavelength of the \alfven wave.

This mean electric field $E_{\parallel}$ is smaller than the (non-dissipative)
spike of $E_\parallel$ that was needed in the two-fluid model to polarize the
background plasma. The fractional dissipation rate of the wave energy density
$U_w$ can be estimated as
\begin{equation}
    \label{eq:diss-relative}
    \frac{\dot{U}_{w}}{U_{w}} \sim \frac{E_{\parallel}j_{A}}{B^{2}_{w}/8\pi} \sim \frac{\sqrt{\kappa}c}{\sigma_{w}\lambda}.
\end{equation}
Thus, a fraction $\sqrt{\kappa}/\sigma_w$ of the initial electromagnetic energy
in the \alfven wave is dissipated in one crossing time of the wave $\lambda/c$.

This slow acceleration should saturate if beam reaches the group speed of the
wave,
\begin{equation}
    v_{A} = \sqrt{\frac{\sigma}{1 + \sigma}}, \quad \gamma_{A} = \sqrt{1+\sigma},
\end{equation}
where $\sigma = B_{0}^{2}/4\pi n_{w}m_{e}c^{2}$ is the magnetization parameter,
and $n_w$ is the plasma density in the wave, which can significantly exceed the
background $n_0$ in the regime of $\kappa \gg 1$.
The acceleration of the plasma beam carried by the wave should saturate when the
beam Lorentz factor reaches $\gamma \sim \sqrt{\sigma}$, and further dissipation
will likely go into heating up the plasma beam. Note, however, that it can take
a long time for the wave to reach this saturated regime, and it may not occur in
a real neutron star magnetosphere.

The fractional dissipation rate \eqref{eq:diss-relative} scales with $\kappa$
and $\sigma_{w}$, and it is independent of the \alfven wave amplitude
$B_w/B_0$. We performed a series of simulations with the same $B_w$ and
different background magnetic field strengths $B_0$, such that $B_{w}/B_{0}$
ranges from 0.01 to 0.5. We find that in all cases, as long as $B_{w}$ and
$\kappa$ remains constant and $\sigma_{w}\gg\sqrt{\kappa}$, the energy
dissipation rates and particle acceleration histories are identical.

Using the volume of the emission region
$V \sim \lambda \ell_{\perp} 2\pi r\sin\theta$, one can estimate the maximum
luminosity from this slow dissipation if the particle kinetic energy gain can be
converted to emission at some wavelengths:
\begin{equation}
\label{eq:Ldiss}
    \begin{split}
        L_\mathrm{diss} &\sim E_{\parallel}j_{A}V \\
        &\sim \frac{\sqrt{\kappa}c}{\lambda}\frac{k_{\perp}B_{w}m_{e}c^{2}}{8\pi e}\lambda \ell_{\perp} 2\pi r\sin\theta \\
        &\sim \sqrt{\kappa}B_{w} r^{3/2}\frac{\pi m_{e}c^{3}}{2e}.
    \end{split}
\end{equation}
Since 
$B_w=B_w^\star(r/r_{*})^{-3/2}$, the dissipation power is essentially only
dependent on $\kappa$ and the initial \alfven wave amplitude emitted by the
star, $B_w^\star$.

\subsection{Wave Propagation through a Density Jump}

\label{sec:jump-n0}

The numerical results described in Section~\ref{sec:const-n0} apply to a wave
undergoing slow shearing, or propagating in a background plasma with a slowly
varying density. We now investigate the opposite regime where $\kappa$ increases
suddenly, on a length scale that is comparable or shorter than the \alfven
wavelength. In particular, we wish to check whether there is any dramatic
transient behavior in the extreme limit when the \alfven wave propagates across
a sharp boundary where $\kappa$ transitions from $<1$ to $\gg 1$.

\begin{figure}[t]
    \centering
    \plotone{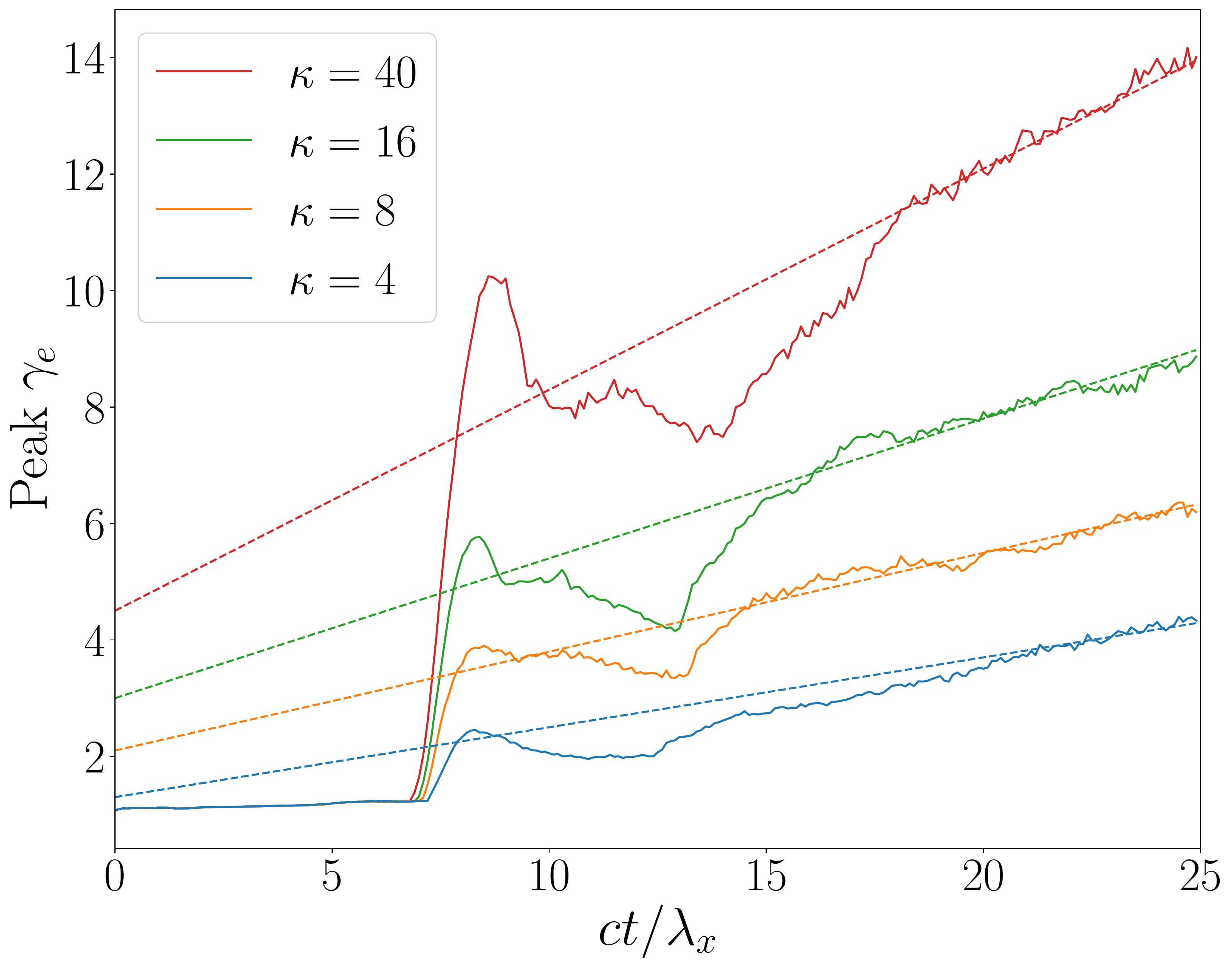}
    \caption{Evolution of maximum electron Lorentz factor vs $\kappa$ when there
      is a density jump. Dashed lines have slopes $0.4\sqrt{\kappa}/\lambda$,
      and are acceleration models with constant $E_{\parallel}$. After the
      transient when the wave goes across the density jump, particle
      acceleration proceeds in a similar fashion as Figure~\ref{fig:kappa}.}
    \label{fig:kappa-jump}
\end{figure}

We use the same simulation setup as described in Section~\ref{sec:const-n0},
with the exception that $n_{0}$ drops sharply at $x = 0.4L_{x}$. We have carried
out a series of simulations where $\kappa = 0.8$ for $x < 0.4L_{x}$ and $\kappa$
for $x > 0.4L_{x}$ is a constant value above unity. In our simulations, the
value of $\kappa$ after the jump ranges from 4 to 40.
Figure~\ref{fig:kappa-jump} shows the evolution of peak electron Lorentz factors
for these runs before and after the density jump.

We find that during its encounter with the density jump, the \alfven wave
induces a coherent $E_{\parallel} \propto \sqrt{\kappa}/\lambda_{x}$ to quickly
accelerate charges of the right sign to $\gamma \sim \sqrt{\kappa}$. The
parallel electric field separates the charges and sweeps the right amount of
$e^{\pm}$ with the wave to conduct the required current. The most significant
acceleration happens near the leading edge of the wave, which is negatively
charged in our wave profile (see panel (c) of Figure~\ref{fig:simulation}). The
encounter phase with the high $E_\parallel$ has a short duration, and the total
dissipated energy during the simulation remains to be dominated by the later
phase, when the wave continues to propagate through the low-density background.
At this late phase, the wave behavior is similar to that found in Section 4.2
with $\kappa=\mathrm{const}$.

\section{Discussion}

We have studied the propagation of \alfven waves in different plasma densities,
in particular when the background plasma density is insufficient to support the
required current $j_A$. In this ``charge-starved'' regime, the \alfven wave
manages to still provide the required current and charge densities by sweeping
the charges of the right sign with it. The wave
becomes charge-separated rather than truly charge-starved, and the
charge carriers move at near the speed of light. In the highly magnetized regime
relevant to neutron star magnetospheres, only a small amount of the \alfven
wave energy needs to be converted to particle kinetic energy to sustain this
configuration.

The particle acceleration process is slow and smooth, with only a small induced
$E_{\parallel}$ on average. There are two contributions to the acceleration of
particles. As the ``charge-starvedness'' parameter $\kappa$ increases, the
characteristic Lorentz factor for the charge carriers in the wave increases as
$\sqrt{\kappa}$. At the same time, when $\kappa \gtrsim 1$, the current carrying
particles are also accelerated by a nonzero dissipative 
$E_{\parallel} \propto \sqrt{\kappa}/\lambda$, even when $\kappa$ is
constant.

We find that the transition to the regime of $\kappa>1$ leads to the dissipation
rate given by the simple estimate~(\ref{eq:Ldiss}). Below we use the parameters
of the X-ray bursts from the galactic magnetar SGR 1935+2154 as an illustrative
example. The energy budget of the X-ray burst is consistent with an \alfven
wave of amplitude $B_w^\star \sim 10^{-3}B_{p}\sim 10^{11}\,\mathrm{G}$
\citep{2020ApJ...900L..21Y}. Assuming a large $\kappa \sim 100$, we can then
estimate the maximum luminosity generated by the wave entering charge starvation
as
\begin{equation}
    \label{eq:Ldiss1}
    L_\mathrm{diss} \approx 8\times 10^{34}\left(\frac{\kappa}{100}\right)^{1/2}
    \left(\frac{
    B_w^\star }
    {10^{11}\,\mathrm{G}}\right)\,\mathrm{erg/s}.
\end{equation}
It is 5 orders of magnitude lower than the X-ray burst luminosity. It is also
much lower than the luminosity of the FRB produced by SGR~1935+2154. The FRB
energy output was $\sim 3\times 10^{34}\,\mathrm{erg}$, lasting about
$1\,\mathrm{ms}$, which implies an isotropic equivalent luminosity of
$\sim 3\times 10^{37}\,\mathrm{erg/s}$ \citep{2020arXiv200510324T}. Even
assuming a huge $\kappa$ and a very high radiation efficiency of order unity,
the luminosity (\ref{eq:Ldiss1}) is insufficient to power the observed FRB.
Furthermore, our simulations do not show strong bunching of $e^{\pm}$ in the
saturated plasma oscillations in the \alfven wave, and therefore we do not
expect efficient coherent emission. Even if the plasma 
did form bunches, the particles do not gain enough energy or sufficiently high
Lorentz factors for coherent emission in the radio band. Therefore, we conclude
that it is unlikely that FRBs are produced through the charge-starvation
mechanism alone.

Our results also imply that an \alfven wave entering charge starvation does not
need to spawn new $e^\pm$ pairs to propagate. Acceleration of particles to
pair-producing energies likely requires additional mechanisms, such as wave
collisions and nonlinear cascades
\citep[e.g.][]{PhysRevD.57.3219,2019ApJ...881...13L}. For instance, these
processes may be essential for the quake-excited \alfven waves during the glitch
chocking of the Vela pulsar radio emission.

In the vicinity of magnetars, resonant inverse-Compton scattering of the thermal
X-ray photons flowing from the star can exert an efficient drag force on the
plasma even when particle Lorentz factors are on the order of $\sim 10$,
depending on the position in the magnetosphere \citep{2013ApJ...777..114B,
  2020arXiv200808659T}. This can create a significant additional channel for
dissipation of the \alfven wave, which can potentially convert most of the
energy gained by the particles into hard X-ray emission. The resulting X-ray
luminosity may well be in the observable range. The inverse Compton scattering
may also induce pair production which increases the background plasma density
and thus reduces $\kappa$. These effects will be studied in a future work.

\bigskip

We thank Dmitri Uzdensky and Jens Mahlmann for helpful discussions. A.C. is
supported by NSF grants AST-1806084 and AST-1903335. Y.Y. is supported by a
Flatiron Research Fellowship at the Flatiron Institute, Simons Foundation.
A.M.B. is supported by NASA grant NNX\,17AK37G, NSF grant AST\,2009453, Simons
Foundation grant \#446228, and the Humboldt Foundation. Research at Perimeter
Institute is supported in part by the Government of Canada through the
Department of Innovation, Science and Economic Development Canada and by the
Province of Ontario through the Ministry of Colleges and Universities.


\begin{thebibliography}{}
\expandafter\ifx\csname natexlab\endcsname\relax\def\natexlab#1{#1}\fi
\providecommand{\url}[1]{\href{#1}{#1}}
\providecommand{\dodoi}[1]{doi:~\href{http://doi.org/#1}{\nolinkurl{#1}}}
\providecommand{\doeprint}[1]{\href{http://ascl.net/#1}{\nolinkurl{http://ascl.net/#1}}}
\providecommand{\doarXiv}[1]{\href{https://arxiv.org/abs/#1}{\nolinkurl{https://arxiv.org/abs/#1}}}

\bibitem[{{Beloborodov}(2013)}]{2013ApJ...777..114B}
{Beloborodov}, A.~M. 2013, \apj, 777, 114, \dodoi{10.1088/0004-637X/777/2/114}

\bibitem[{{Blaes} {et~al.}(1989){Blaes}, {Blandford}, {Goldreich}, \&
  {Madau}}]{1989ApJ...343..839B}
{Blaes}, O., {Blandford}, R., {Goldreich}, P., \& {Madau}, P. 1989, \apj, 343,
  839, \dodoi{10.1086/167754}

\bibitem[{{Bochenek} {et~al.}(2020){Bochenek}, {Ravi}, {Belov}, {Hallinan},
  {Kocz}, {Kulkarni}, \& {McKenna}}]{2020arXiv200510828B}
{Bochenek}, C.~D., {Ravi}, V., {Belov}, K.~V., {et~al.} 2020, arXiv e-prints,
  arXiv:2005.10828.
\newblock \doarXiv{2005.10828}

\bibitem[{{Bransgrove} {et~al.}(2020){Bransgrove}, {Beloborodov}, \&
  {Levin}}]{2020ApJ...897..173B}
{Bransgrove}, A., {Beloborodov}, A.~M., \& {Levin}, Y. 2020, {A Quake Quenching
  the Vela Pulsar}, \dodoi{10.3847/1538-4357/ab93b7}

\bibitem[{{Duncan} \& {Thompson}(1992)}]{1992ApJ...392L...9D}
{Duncan}, R.~C., \& {Thompson}, C. 1992, \apjl, 392, L9, \dodoi{10.1086/186413}

\bibitem[{{Goldreich} \& {Julian}(1969)}]{1969ApJ...157..869G}
{Goldreich}, P., \& {Julian}, W.~H. 1969, \apj, 157, 869,
  \dodoi{10.1086/150119}

\bibitem[{{Kumar} \& {Bo{\v{s}}njak}(2020)}]{2020MNRAS.494.2385K}
{Kumar}, P., \& {Bo{\v{s}}njak}, {\v{Z}}. 2020, \mnras, 494, 2385,
  \dodoi{10.1093/mnras/staa774}

\bibitem[{{Kumar} {et~al.}(2017){Kumar}, {Lu}, \&
  {Bhattacharya}}]{2017MNRAS.468.2726K}
{Kumar}, P., {Lu}, W., \& {Bhattacharya}, M. 2017, \mnras, 468, 2726,
  \dodoi{10.1093/mnras/stx665}

\bibitem[{{Li} {et~al.}(2019){Li}, {Zrake}, \&
  {Beloborodov}}]{2019ApJ...881...13L}
{Li}, X., {Zrake}, J., \& {Beloborodov}, A.~M. 2019, \apj, 881, 13,
  \dodoi{10.3847/1538-4357/ab2a03}

\bibitem[{{Lu} {et~al.}(2020){Lu}, {Kumar}, \& {Zhang}}]{2020arXiv200506736L}
{Lu}, W., {Kumar}, P., \& {Zhang}, B. 2020, arXiv e-prints, arXiv:2005.06736.
\newblock \doarXiv{2005.06736}

\bibitem[{{Melrose}(1986)}]{1986islp.book.....M}
{Melrose}, D.~B. 1986, {Instabilities in Space and Laboratory Plasmas}

\bibitem[{{Mereghetti} {et~al.}(2020){Mereghetti}, {Savchenko}, {Ferrigno},
  {G{\"o}tz}, {Rigoselli}, {Tiengo}, {Bazzano}, {Bozzo}, {Coleiro},
  {Courvoisier}, {Doyle}, {Goldwurm}, {Hanlon}, {Jourdain}, {von Kienlin},
  {Lutovinov}, {Martin-Carrillo}, {Molkov}, {Natalucci}, {Onori}, {Panessa},
  {Rodi}, {Rodriguez}, {S{\'a}nchez-Fern{\'a}ndez}, {Sunyaev}, \&
  {Ubertini}}]{2020arXiv200506335M}
{Mereghetti}, S., {Savchenko}, V., {Ferrigno}, C., {et~al.} 2020, arXiv
  e-prints, arXiv:2005.06335.
\newblock \doarXiv{2005.06335}

\bibitem[{{Palfreyman} {et~al.}(2018){Palfreyman}, {Dickey}, {Hotan},
  {Ellingsen}, \& {van Straten}}]{2018Natur.556..219P}
{Palfreyman}, J., {Dickey}, J.~M., {Hotan}, A., {Ellingsen}, S., \& {van
  Straten}, W. 2018, \nat, 556, 219, \dodoi{10.1038/s41586-018-0001-x}

\bibitem[{{Ruderman}(1976)}]{1976ApJ...203..213R}
{Ruderman}, M. 1976, \apj, 203, 213, \dodoi{10.1086/154069}

\bibitem[{{Swisdak}(2006)}]{2006physics...6044S}
{Swisdak}, M. 2006, arXiv e-prints, physics/0606044.
\newblock \doarXiv{physics/0606044}

\bibitem[{{The CHIME/FRB Collaboration} {et~al.}(2020){The CHIME/FRB
  Collaboration}, {:}, {Andersen}, {Band ura}, {Bhardwaj}, {Bij}, {Boyce},
  {Boyle}, {Brar}, {Cassanelli}, {Chawla}, {Chen}, {Cliche}, {Cook},
  {Cubranic}, {Curtin}, {Denman}, {Dobbs}, {Dong}, {Fandino}, {Fonseca},
  {Gaensler}, {Giri}, {Good}, {Halpern}, {Hill}, {Hinshaw}, {H{\"o}fer},
  {Josephy}, {Kania}, {Kaspi}, {Landecker}, {Leung}, {Li}, {Lin}, {Masui},
  {Mckinven}, {Mena-Parra}, {Merryfield}, {Meyers}, {Michilli}, {Milutinovic},
  {Mirhosseini}, {M{\"u}nchmeyer}, {Naidu}, {Newburgh}, {Ng}, {Patel}, {Pen},
  {Pinsonneault-Marotte}, {Pleunis}, {Quine}, {Rafiei-Ravandi}, {Rahman},
  {Ransom}, {Renard}, {Sanghavi}, {Scholz}, {Shaw}, {Shin}, {Siegel}, {Singh},
  {Smegal}, {Smith}, {Stairs}, {Tan}, {Tendulkar}, {Tretyakov}, {Vanderlinde},
  {Wang}, {Wulf}, \& {Zwaniga}}]{2020arXiv200510324T}
{The CHIME/FRB Collaboration}, {:}, {Andersen}, B.~C., {et~al.} 2020, arXiv
  e-prints, arXiv:2005.10324.
\newblock \doarXiv{2005.10324}

\bibitem[{Thompson \& Blaes(1998)}]{PhysRevD.57.3219}
Thompson, C., \& Blaes, O. 1998, Phys. Rev. D, 57, 3219,
  \dodoi{10.1103/PhysRevD.57.3219}

\bibitem[{{Thompson} \& {Kostenko}(2020)}]{2020arXiv200808659T}
{Thompson}, C., \& {Kostenko}, A. 2020, arXiv e-prints, arXiv:2008.08659.
\newblock \doarXiv{2008.08659}

\bibitem[{{Yuan} {et~al.}(2020){Yuan}, {Beloborodov}, {Chen}, \&
  {Levin}}]{2020ApJ...900L..21Y}
{Yuan}, Y., {Beloborodov}, A.~M., {Chen}, A.~Y., \& {Levin}, Y. 2020, \apjl,
  900, L21, \dodoi{10.3847/2041-8213/abafa8}

\end{thebibliography}
\bibliographystyle{aasjournal}

\end{document}